%% file: kochukhov.tex
\documentclass[mypaper,8pt,twoside]{CoAst}
\usepackage{epsf,graphicx,fancyhdr}
\input{CoAst_layo}

\newcommand{\ms}{m\,s$^{-1}$}
\newcommand{\pr}{Pr~{\sc iii}}
\newcommand{\nd}{Nd~{\sc iii}}

\newcommand{\fifps}[2]{\centering\resizebox{#1}{!}{\includegraphics{#2}}}

\newcommand{\fizps}[2]{\centering\resizebox{#1}{!}{\rotatebox{270}{\includegraphics{#2}}}}

\begin{document}
\sf

\chapterDSSN{Observations of pulsations in roAp stars}{O. Kochukhov}

\Authors{O. Kochukhov} 
\Address{Department of Astronomy and Space Physics, Uppsala University, \\
Box 515, SE-751 20 Uppsala, Sweden}

\noindent
\begin{abstract}
I review recent results of the observations of magnetoacoustic {\it p-}mode
oscillations in roAp stars with the focus on time-resolved spectroscopic studies.
Time-series spectroscopy of roAp stars reveal unexpected and diverse pulsational
behaviour in the spectral lines of different chemical elements. These unique pulsational
characteristics arise from an interplay between short vertical
length of pulsation waves and extreme chemical stratification in the 
atmospheres of peculiar stars. This enables a tomographic reconstruction of the
depth-dependence of chemical composition and pulsation wave properties.
Combination of magnetoacoustic tomography with the Doppler imaging of the horizontal
non-radial pulsation pattern opens possibility for an unprecedented three-dimensional mapping of 
roAp atmospheres.
\end{abstract}

\section{Introduction}

Significant fraction of the upper main sequence stars of spectral classes between B and F
possesses strong, ordered magnetic field and shows surface chemical composition strongly deviating
from that of the Sun. These chemically peculiar Ap/Bp stars are characterized
by an unusually slow rotation and show spectacular variability of the field strength, mean
brightness and spectra on the rotation time scale. This is understood to be a result of the
rotational modulation of the aspect at which stable stellar magnetic field geometry and surface
chemical inhomogeneities are observed.

In addition to their remarkable magnetic and chemical surface characteristics, many cooler
Ap stars also exhibit high-overtone non-radial acoustic {\it p-}modes. There are more than
30 such rapidly oscillating Ap (roAp) stars known at the present time (Kurtz \& Martinez
2000). These objects oscillate with periods in the range of 6--21~min, 
while their light variation amplitudes rarely exceed 10~mmag in Johnson $B$. 
Photometric investigations of roAp stars carried out during the last 25 years have yielded
unique asteroseismic information on the internal structure and fundamental parameters of
roAp pulsators (e.g., Matthews et al. 1999, Cunha et al. 2003).

The observed pulsation amplitudes of roAp stars are modulated according to the visible
magnetic field structure, pointing to a defining role played by magnetic fields in exciting
the oscillations and shaping the main pulsation properties. Observation of the coincidence
of the magnetic field and pulsation amplitude extrema gave rise to the {\it oblique pulsator
model} (OPM, Kurtz 1982), which attribites  the main characteristics of roAp pulsations to
an oblique $\ell=1$, $m=0$ mode, aligned with the axis of a  quasi-dipolar magnetic field.
The OPM gave rather successful geometrical explanation of the main features in the roAp
frequency spectra. However, subsequent detailed studies of roAp pulsations have revealed
that the mode geometry in some stars defies a simple interpretation in terms of a single
spherical harmonic (e.g., Kurtz et al. 1997). 

Several theoretical investigations (Bigot \& Dziembowski 2002, Saio \& Gautschy 2004, Saio
2005) studied the effects of the distortion of oblique pulsation mode geometry by the global
magnetic field and stellar rotation. Bigot \& Dziembowski (2002) suggested that the
rotational distortion of pulsation eigenmodes is represented by a superposition of
non-axisymmetric spherical harmonic components, and that there is no alignment of the
pulsation axis and the dipolar magnetic field. On the other hand, Saio \& Gautschy (2004)
and Saio (2005) found an axisymmetric pulsation structure aligned with the magnetic field
and predicted that the $\ell=1$ mode should be distorted by a dipolar magnetic field in such
a way that pulsation amplitude is strongly confined to the magnetic axis.

Sophistication of these theoretical models notwithstanding, it became clear that modelling
of the photometric light curves of roAp stars is  unable to provide useful tests of
magnetoacoustic theories. The information content of the time-resolved photometric
observations is small due to averaging of the pulsational disturbances over the visible
stellar hemisphere and is also highly uncertain because rapid light variation in roAp stars
involves non-linear and non-adiabatic effects that are poorly understood (Medupe \& Kurtz
1998). The latter problem explains why no consistent physical picture of the photometric
variability of roAp has ever been developed. Instead of deducing the structure of the
luminosity perturbations from first principles, all attempts to interpret photometric
observations of roAp stars have {\it assumed} that luminosity perturbations are proportional
to pulsational displacement. Therefore, constraints on the pulsation mode geometry obtained
from the photometry of roAp stars are inherently indirect, which arguably makes any
subsequent inferences about the physics of magnetoacoustic oscillations questionable.

\section{Spectroscopic studies of roAp pulsations}

Investigation of pulsational variation in the spectral line profiles of roAp stars
observed at high time and spectral resolution provides a much more direct and
unprecedentedly rich information about the vertical and horizontal structure of {\it
p-}modes and about their relation to the magnetic field topology, chemical inhomogeneities
and anomalous atmospheric structure of Ap stars. 

Recently major progress in the observational study of roAp stars was achieved by employing
time-series spectroscopy. Time-resolved observations of magnetic pulsators revealed a
surprising diversity, not observed in any other type of pulsating stars, in oscillations of
different lines (e.g., Kanaan \& Hatzes 1998). Detailed analysis of the bright roAp star
$\gamma$~Equ (Kochukhov \& Ryabchikova 2001a) demonstrated that spectroscopic pulsational
variability is dominated by the lines of rare-earth ions, especially those of Pr and Nd. On
the other hand, light and iron-peak elements do not pulsate with amplitudes above
50--100~\ms. This is at least an order of magnitude lower than the $\sim$\,1~\kms\
variability observed in the lines of rare-earth elements (REE). Many other roAp stars have
been found to show a very similar overall pulsational behaviour (Kochukhov \& Ryabchikova
2001b, Balona 2002, Mkrtichian et al. 2003, Ryabchikova et al. 2006).

\section{Magnetoacoustic tomography}

The peculiar characteristics of the {\it p-}mode pulsations in roAp stars were clarified by
Ryabchikova et al. (2002), who were the first to relate pulsational variability to vertical
stratification of chemical elements. This study of the atmospheric properties of 
$\gamma$~Equ showed that the light and iron-peak elements are enhanced in the lower
atmospheric layers ($\log\tau_{5000}\ge-0.5$), whereas REE ions are concentrated in a cloud
with a lower boundary at $\log\tau_{5000}\le-4$ (Mashonkina et al. 2005). Thus,
high-amplitude pulsations observed in REE lines occur in the upper atmosphere, while lines
of elements showing no significant variability form in the lower atmosphere. This leads to
the following general picture of roAp pulsations: we observe a signature of a
magnetoacoustic wave, propagating outwards with increasing amplitude through the chemically
stratified atmosphere. 

The presence of significant phase shifts between the pulsation radial velocity curves of
REEs (Kochukhov \& Ryabchikova 2001a), or even between lines of the same element
(Mkrtichian et al. 2003), can be attributed to the chemical stratification effects
and, possibly, to the short vertical wavelength of the running magnetoacoustic wave. These
unique properties of roAp pulsations, combined with a presence of large vertical abundance
gradients in the line-forming region, make it possible to resolve the vertical structure of
{\it p-}modes and to study propagation of pulsation waves at the level of detail previously
possible only for the Sun.

The study by Ryabchikova et al. (2002) represents the first attempt to use the vertical
chemical inhomogeneities as spatial filters which resolve the vertical {\it p-}mode
structure. The basic idea of this {\it pulsation tomography} approach consists of
characterizing pulsational behaviour of a sample of spectral lines and subsequent
interpretation of these observations in terms of the pulsation wave propagation. Chemical
stratification analysis of REE lines constrains formation depths of pulsating lines, thus
allowing one to associate geometrical height with the amplitude and phase of RV pulsations.

\begin{figure}[!t]
\fizps{5.5cm}{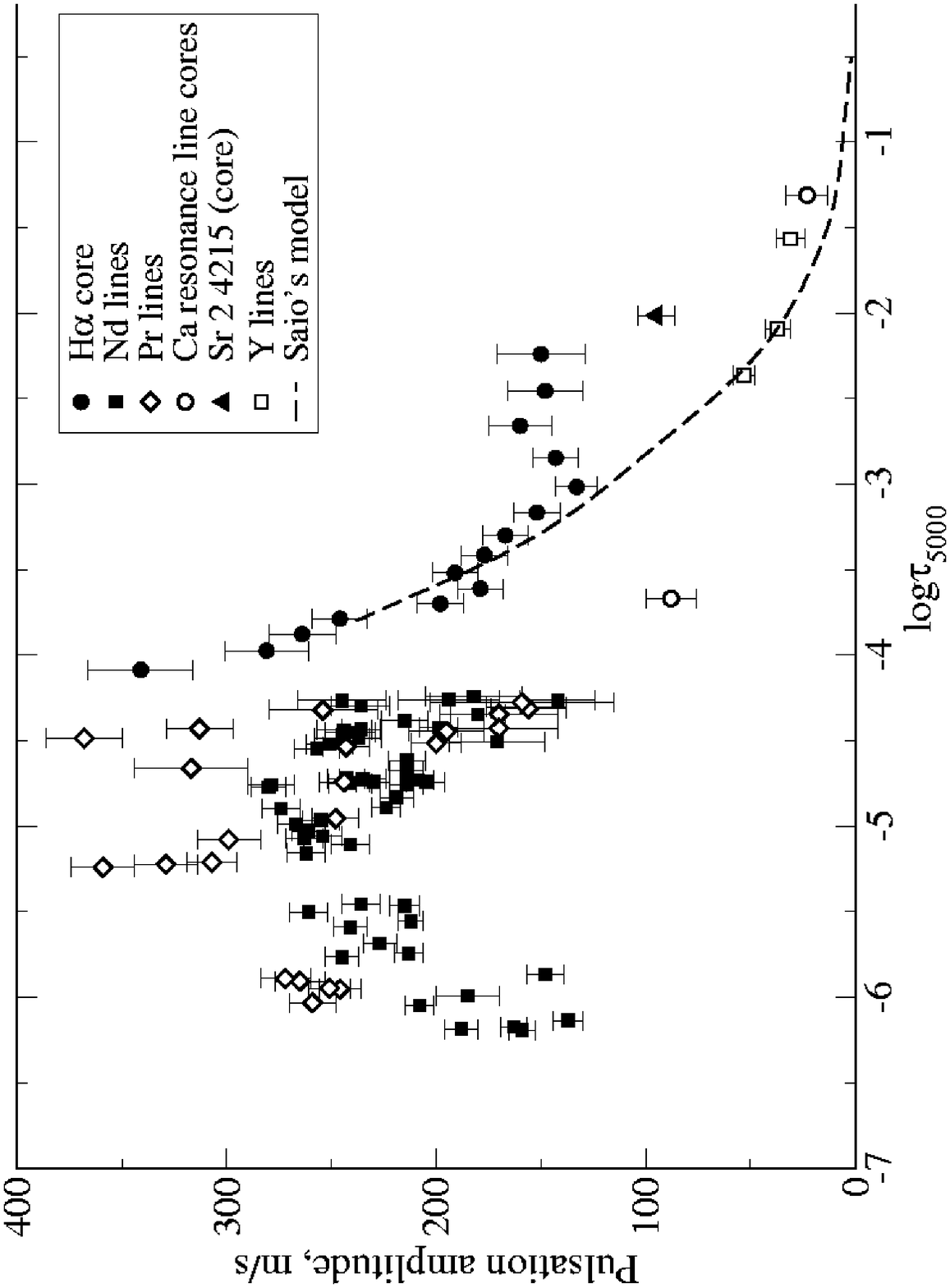}
\fizps{5.5cm}{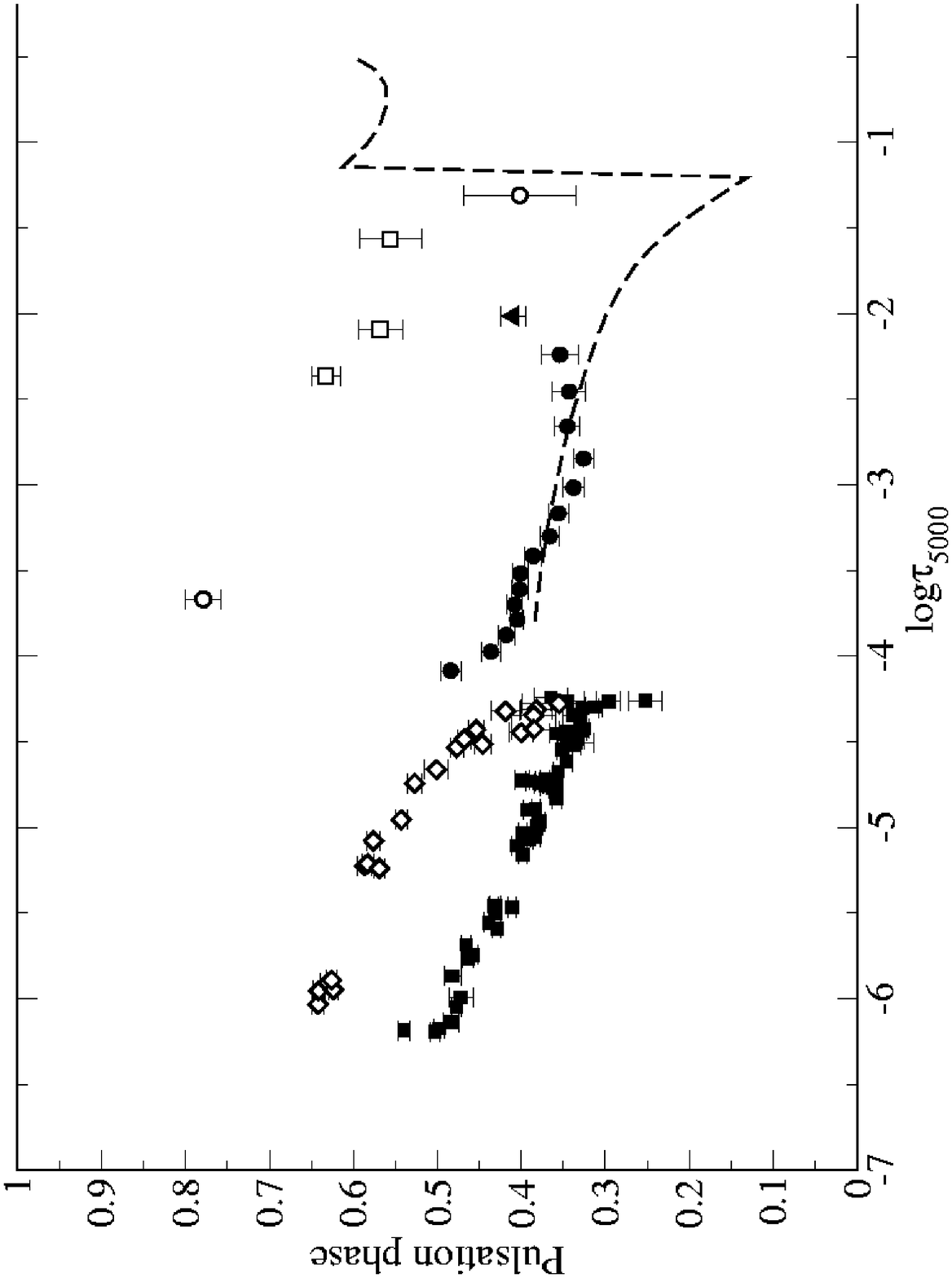}
\caption{Reconstruction of the vertical cross-section of pulsation mode in 
the roAp star HD\,24712. Symbols show the observed amplitude
(left panel) and phase (right panel) of the radial velocity variation for different 
spectral lines. The dashed line illustrates theoretical depth dependence of the pulsation
wave properties (H. Saio, private communication).
}
\label{fig:hd24712}
\end{figure}

Fig.~\ref{fig:hd24712} illustrates results of the pulsation tomography analysis of the roAp
star HD\,24712. This star was observed simultaneously by the MOST satellite and from the
ground, using high-resolution spectrographs at several large telescopes, including the ESO
VLT. Using these time-series spectra, Ryabchikova et al. (2006) studied pulsational
variation of more than 600 lines. Pulsation amplitudes and phases for several characteristic
lines of light elements, the core of H$\alpha$ and numerous REE lines are plotted in
Fig.~\ref{fig:hd24712} as a function of optical depth. Observations are compared with the
theoretical {\it p-}mode cross-section (H. Saio, private communication). NLTE line formation
was taken into account in chemical stratification analysis of REE ions. This modelling
reveals a rapid increase of the pulsation amplitude with height and the respective change
of the pulsation phase. Oscillation amplitude reaches maximum at
$\log\tau_{5000}\approx-4.5$ and decreases in the higher layers. 

Preliminary NLTE stratification analysis of Pr (Ryabchikova et al., this conference)
suggests that formation heights of the \pr\ absorption features are not too different
from \nd, despite a clear phase offset between the two groups of lines in HD\,24712. This
phase difference, as well as the amplitude and phase jump  between the uppermost layers
probed by the H$\alpha$ core and the location of the REE cloud, may reflect shortcomings
of the complicated NLTE analysis. Alternatively, it is possible that we are seeing
effects of the inhomogeneous surface distribution of different REEs in HD\,24712.
Magnetic Doppler images obtained by L\"uftinger et al. (this conference) show that the
horizontal geometry of the Pr and Nd distribution is not the same. This may lead to
different pulsational behaviour because the vertical structure of magnetoacoustic modes
depends on the field strength and orientation (Cunha 2006) and, therefore, may be
different at the locations of the Pr and Nd spots in HD\,24712.

The detailed pulsation tomography analysis based on the NLTE chemical stratification
modelling is very demanding in terms of the quality of observations, required input data and
computer resources. This is why only two roAp stars, $\gamma$~Equ and HD\,24712, were
studied with this method up to now. A different approach to the pulsation tomography problem was
proposed by Ryabchikova et al. (2007). They noted that in the framework of the
outward propagating magnetoacoustic wave one expects a continuous amplitude vs.
phase relation for pulsation modes. The amplitude-phase diagrams offer a possibility
to trace the vertical variation in the  mode structure without assigning physical depth to
pulsation measurements. Ryabchikova et al. (2007) analysed a sample of ten roAp stars,
measuring variation of several hundred lines for each object. The amplitude-phase diagrams
were constructed for each star and the resultant vertical mode cross-sections were compared
with other pulsational characteristics and with the fundamental stellar parameters. As an
outcome of this analysis, it was discovered that the form of pulsational perturbation
changes from predominantly standing to mainly running wave within the REE line-forming
region. It appears that the location of this interesting modification of the pulsation wave
properties shifts towards higher layers for cooler roAp stars.

\section{Variability of line bisectors}

In addition to the diversity in the pulsation signatures of different elements and ions,
variation of individual strong REE lines in roAp stars is far from trivial. The most
surprising observation is a large change in the amplitude and phase of bisector RV  with
intensity inside individual lines. In some sharp-lined roAp stars RV amplitude increases
from a few hundred \ms\ to several \kms\ as one moves towards the outer parts of the
line  profiles. At the same time, pulsation phase shows complicated trends with bisector
intensity, sometimes changing by up to 180$^{\rm o}$. This remarkable bisector variation
was first discovered in $\gamma$~Equ (Sachkov et al. 2004) and has been demonstrated for
other roAp stars (Kurtz et al. 2005, Ryabchikova et al. 2006, 2007). The rapidly rotating
pulsators, such as HD\,99563 (Elkin et al. 2005), show increase of bisector amplitude
towards the core -- a trend opposite to that of $\gamma$~Equ.

The line core and wings are expected to sample somewhat different parts of the atmosphere
and this why changes in the bisector variation across individual lines are often  attributed
to the  height effects. According to this explanation, modification of bisector oscillations
indicates a remarkably complex and rapidly changing pulsation mode structure, with several
nodes located in the line-forming region. However, this interpretation must be viewed with
caution. A rapid change of the pulsation wave properties with height is inconsistent with
the well-defined, smooth amplitude and phase  depth dependence inferred by the pulsation
tomography. Theoretical models (Saio \& Gaustchy 2004, Saio 2005) also predict no nodes in
the upper atmosphere. Moreover, an implicit assumption that any deviation from a constant
amplitude and phase of the bisector should be interpreted as a height effect is
questionable. In fact, no studies looked at the bisector behaviour in normal non-radial
pulsators. It appears that at least part of the core-to-wing change of bisector amplitude
may be ascribed to the presence of high-$\ell$ harmonic contribution in the horizontal
pulsation structure. 

Interpretation of the bisector variability is even more ambiguous for rapidly rotating roAp
stars. In these objects the oblique non-radial pulsations are superimposed onto the much
larger velocity field due to the stellar rotation. The primary consequence of the dominant
rotational broadening of spectral lines is that the mapping between bisector intensity and
atmospheric height is no longer valid. Instead, the spectral line wings are formed close to
the limb of the visible stellar disk, whereas the line core region is primarily sensitive to
the disk center. Detailed spectrum synthesis calculations demonstrate that in the presence
of chemical spots one can easily obtain substantial core-to-wing changes of the bisector
amplitude and phase without any depth dependence of pulsation wave. Therefore,
interpretation of the bisector variation of rapidly rotating roAp stars in terms of vertical
structure of {\it p-}modes (e.g., Elkin et al. 2005) is probably  incorrect.

\section{Pulsation Doppler imaging}

The outstanding pulsational variability of REE lines in rapidly rotating roAp
stars permit detailed mapping of the horizontal structure of pulsations. The  oblique nature
of non-radial oscillations allows pulsational monitoring from different aspect
angles, thus facilitating reconstruction of the pulsation pattern. Using this unique
geometrical property of roAp pulsations, Kochukhov (2006) has carried out a high-resolution
spectroscopic monitoring of the prototype roAp star HD\,83368 (HR\,3831). 
This star was observed at the ESO 3.6-m telescope during 6 nights over the period of two weeks. 
Full rotational phase coverage with the time-resolved spectra was obtained, supplying
observational material for the first comprehensive investigation of the 
pulsational line profile variability (LPV) in a roAp star.

The moderately rapid ($v_{\rm e}\sin i=33$~\kms) rotation of HD\,83368 allows to
use the Doppler effect in spectral lines to resolve both the horizontal topology of chemical
inhomogeneities and velocity perturbations due to non-radial oscillations. Kochukhov (2004a)
extended the principles of Doppler imaging (DI) to the reconstruction of the time-dependent
velocity field. In this approach the surface pulsation velocity amplitudes are
recovered directly from the observed line profile variability, without {\it a priori}
constraints on the functional form of pulsation maps. This makes pulsation DI
one of few tools suitable for addressing the daring task of inferring non-radial 
pulsation pattern distorted by the magnetic field and stellar rotation.

Applying pulsation mapping technique to the roAp star HD\,83368, Kochukhov (2004b) obtained
the first stellar Doppler image of the velocity field. DI
analysis of HD\,83368 revealed a nearly axisymmetric pulsation geometry and  for the
first time independently confirmed the alignment of non-radial
pulsation and magnetic field. 
Pulsation mapping finds the oblique pulsator geometry as a
result of the assumption-free analysis, in contrast to all previous studies of roAp stars
which started from the {\it assumption} that OPM is valid.
High-resolution map of pulsations in HD\,83368 were used by Kochukhov (2004b) to
disentangle different harmonic contributions to the pulsation geometry. It was shown that
oscillations are shaped as suggested by Saio \& Gautschy
(2004), whereas the non-axisymmetric pulsation components predicted by the theory of Bigot
\& Dziembowski (2002) cannot be detected. This demonstrates a dominant role of the magnetic
perturbation of the {\it p-}modes and considerably less important influence of the stellar
rotation. 

\section{Rapid line profile variation in sharp-lined roAp stars}

\begin{figure}
\fifps{11cm}{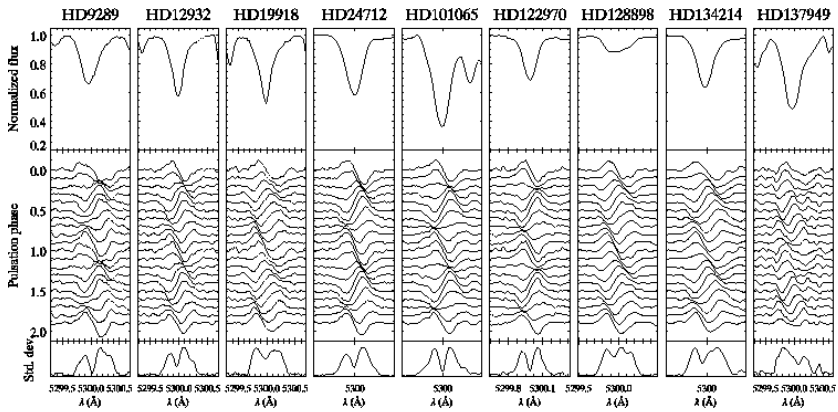}
\caption{Profile variations of the \pr\ 5300~\AA\ line in the spectra of sharp-lined
roAp stars. The average spectrum is plotted in the upper part of each panel.
The time-series of the difference spectra is shown in the middle. The bottom
curve presents the wavelength dependence of the standard deviation.} 
\label{fig:lpv_obs}
\end{figure}

Despite dramatic progress in understanding the vertical structure of pulsation modes in
slowly rotating roAp stars, relatively little attention has been paid to the problem of
inferring the horizontal geometry of pulsations. It is often assumed that a horizontal
cross-section of non-radial pulsation is given by an oblique axisymmetric mode of low
degree, similar to the pulsation geometries found for rapidly rotating roAp stars. Thus,
the question of systematic mode identification has not been thoroughly investigated in the
case of sharp-lined magnetic pulsators, which represent the majority of roAp stars.

Understanding rapid LPV of slowly rotating roAp stars turns out to be a challenging task.
The first observation of roAp line profile variability (Kochukhov \& Ryabchikova 2001a)
demonstrated the presence of unusual blue-to-red running features in the residual spectra of
$\gamma$~Equ. Moreover, a single-wave variability of the REE line width in this star is
clearly inconsistent with {\it any} axisymmetric pulsation geometry described by spherical
harmonics (Aerts et al. 1992, Kochukhov 2005). This led Kochukhov \& Ryabchikova (2001a) to speculate about
the possible presence of non-axisymmetric modes in $\gamma$~Equ -- a suggestion equivalent
to stating that the classical OPM is not applicable to this star. Later Shibahashi et al.
(2004) argued that the blue-to-red running waves in the residual spectra of $\gamma$~Equ are
inconsistent with spectral variability expected for any, axisymmetric and non-axisymmetric
alike, low-degree modes.

The puzzle of the pulsational LPV in sharp-lined roAp stars has been solved by Kochukhov et
al. (2007). This study presented a comprehensive survey of profile variability in ten roAp
stars using observations obtained at the VLT and CFHT telescopes. Variation of the REE lines
was investigated in detail and a prominent change of the profile variability pattern with
height was discovered for all roAp stars. Profile variability of at least one rare-earth ion
in each investigated star is characterized by the blue-to-red moving features, previously
discovered in $\gamma$~Equ. Fig.~\ref{fig:lpv_obs} shows example of this interesting
behaviour, common in rapidly rotating non-radial pulsators but completely inexplicable in
the framework of the standard OPM of slowly rotating roAp stars.

\begin{figure}[!t]
%\fifps{5.6cm}{kochukhov_fig3a.eps}\hspace{0.1cm}
%\fifps{5.6cm}{kochukhov_fig3b.eps}
\fifps{8cm}{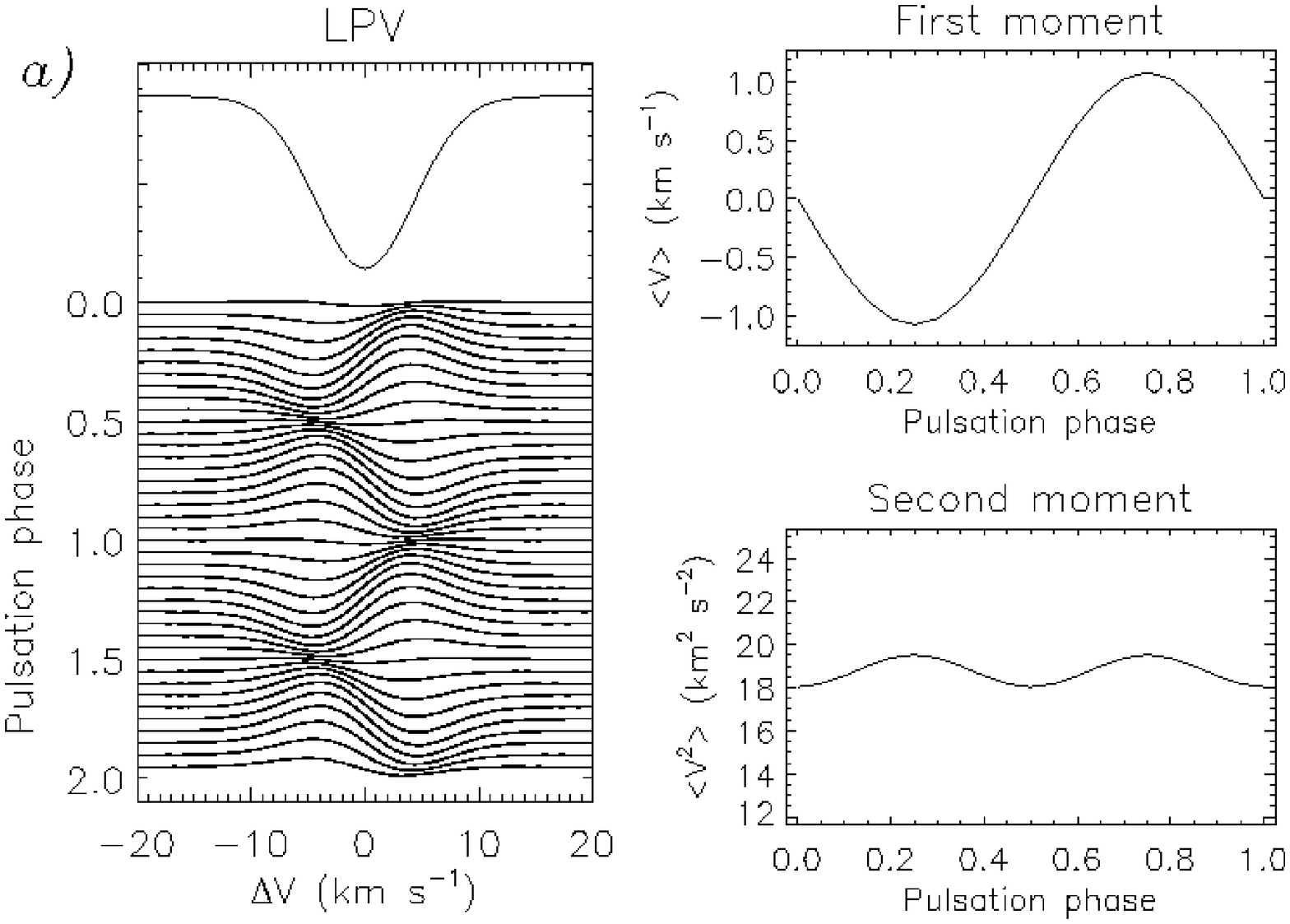}
\fifps{8cm}{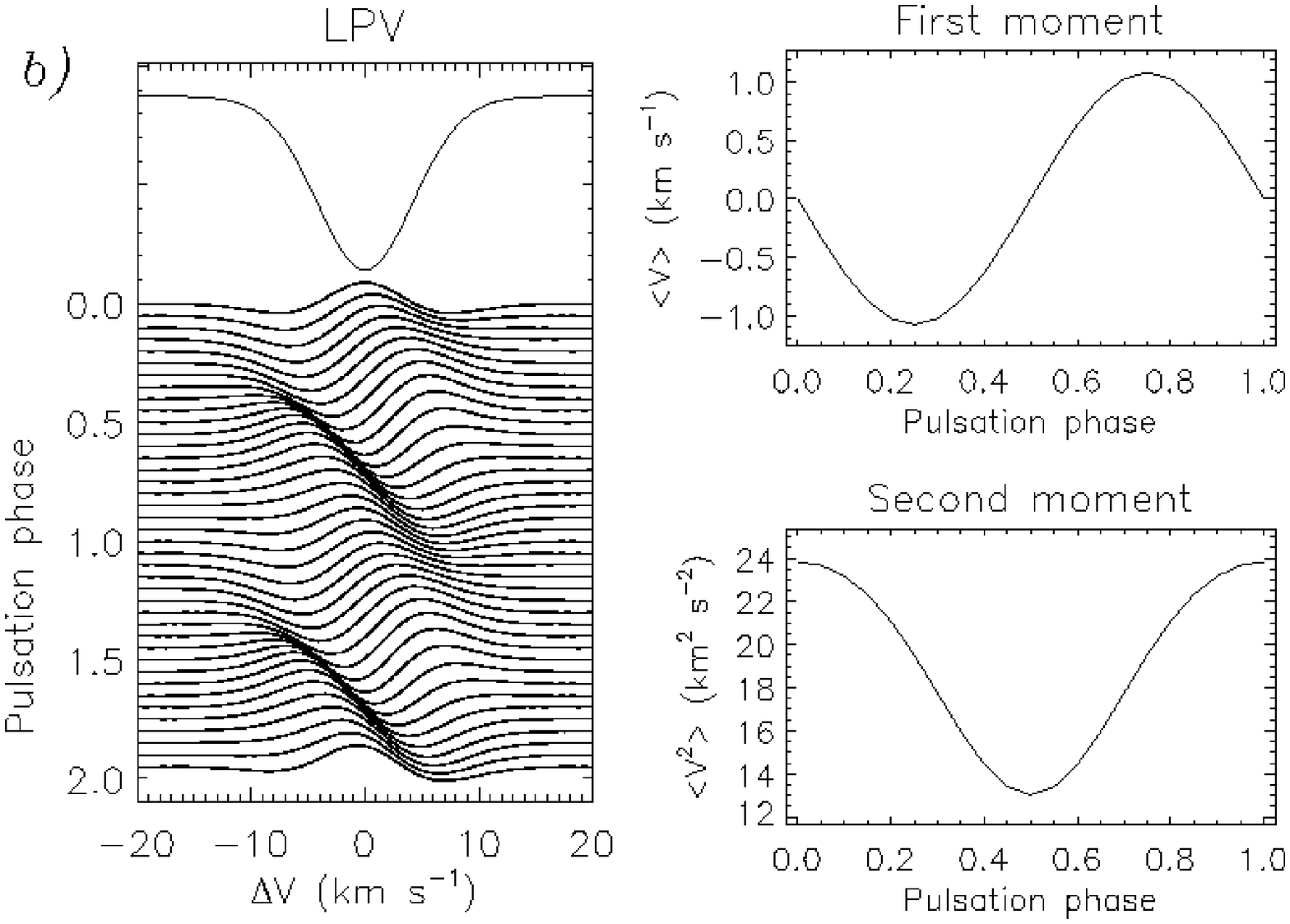}
\caption{Line profile variation of oblique non-radial pulsator. {\bf a)} Spectrum
variability for the $\ell=1$, $m=0$ mode viewed from the pulsation pole. 
{\bf b)} Effect of adding harmonic variability of the line width. 
In each panel the left plot shows the average line profile on
top and time series of the difference spectra below. The right
panels illustrate variation of the first (RV, upper plot) and second 
(line width, lower plot) line profile moments.} 
\label{fig:lpv_mod}
\end{figure}

Analysis of the line profile moments and spectrum synthesis calculations presented by
Kochukhov et al. (2007) demonstrates that unusual oscillations in spectral lines of roAp
stars arise from the pulsational modulation of line widths. This variation occurs
approximately in quadrature with the radial velocity changes, and its amplitude rapidly
increases with height in stellar atmosphere. Kochukhov et al. (2007) proposed that the line
width modulation is a consequence of the periodic expansion and compression of turbulent
layers in the upper atmospheres of roAp stars. This means that the line profile changes
observed in slowly rotating magnetic pulsators should be interpreted as a superposition of
two types of variability: the usual time-dependent velocity field due to an oblique
low-order pulsation mode and an additional line width modulation, synchronized with the
changes of stellar radius. Fig.~\ref{fig:lpv_mod} shows that this new OPM correctly
reproduces the main features in the observed pulsational variability of line profiles and
moments in roAp stars. 

\section{Conclusions and outlook}

Recent investigations of the spectroscopic variability of roAp stars revealed an
interesting and complex picture. The most prominent effect, distinguishing roAp stars
from all other pulsators, is the close interrelation between the chemical stratification
and pulsational variability. Magnetoacoustic waves pass through distinct chemical clouds
in the upper atmospheric layers, giving rise to depth-dependent  amplitude and phase
shifts in the radial velocity variation of different elements.

The quality of observational material and available computing resources have reached the
stage when it becomes feasible to address the task of constructing 3-D dynamical models
of pulsating stellar atmospheres. The pulsation Doppler imaging and magnetoacoustic
tomography techniques can be combined in a self-consistent remote sensing procedure, aimed
at the recovery of 3-D geometry of pulsations and chemical inhomogeneities. This can be
achieved by applying the pulsation Doppler imaging method to the time-series
observations of spectral lines formed at different heights and then combining the
resulting horizontal slices of the pulsation pattern into a 3-D velocity field map.

Construction of the empirical maps should be supported by the advanced theoretical
modelling of peculiar-star atmospheres and pulsations. In particular, a realistic study
of the interaction between pulsations, turbulence and magnetic field in the teneous outer
layers of roAp atmospheres is urgently needed to clarify many puzzling aspects of the
spectroscopic variation of roAp stars.

\acknowledgments{I thank organizers of the
{\it Vienna Workshop on the Future of Asteroseismology} for inviting me to present this
review.
My participation in the workshop was supported by the grants from the Swedish Kungliga 
Fysiografiska S\"allskapet and the Royal Academy of Sciences.}

\References{
Aerts, C., De Pauw, M., Waelkens, C. 1992, \aap, 266, 294\\
Balona, L. A. 2002, \mnras, 337, 1059\\
Bigot, L., Dziembowski, W. A. 2002, \aap, 391, 235\\
Cunha, M. S. 2006, \mnras, 365, 153\\
Cunha, M. S., Fernandes, J. M. M. B., Monteiro, M. J. P. F. G. 2003, \mnras, 343, 831\\
Elkin, V. G., Kurtz, D. W., Mathys, G. 2005, \mnras, 364, 864\\
Kanaan, A., Hatzes, A. P. 1998, \apj, 503, 848\\
Kochukhov, O. 2004a, \aap, 423, 613\\
Kochukhov, O. 2004b, \apj, 615, L149\\
Kochukhov, O. 2005, \aap, 438, 219\\
Kochukhov, O., 2006, \aap, 446, 1051\\
Kochukhov, O., Ryabchikova, T. 2001a, \aap, 374, 615\\
Kochukhov, O., Ryabchikova, T. 2001b, \aap, 377, L22\\
Kochukhov, O., et al. 2007, \mnras, submitted\\
Kurtz, D. W. 1982, \mnras, 200, 807\\
Kurtz, D. W., et al. 1997, \mnras, 287, 69\\
Kurtz, D. W., Martinez, P. 2000, Baltic Astron., 9, 253\\
Kurtz, D. W., Elkin, V. G., Mathys, G. 2005, \mnras, 358, L6\\
Mashonkina, L., Ryabchikova, T., Ryabtsev, A. 2005, \aap, 441, 309\\
Matthews, J. M., Kurtz, D. W., Martinez, P. 1999, \apj, 511, 422\\
Medupe, R., Kurtz, D. W. 1998, \mnras, 299, 371\\
Mkrtichian, D. E., Hatzes, A. P., Kanaan, A. 2003, \mnras, 345, 781\\
Ryabchikova, T., et al. 2002, \aap, 384, 545\\
Ryabchikova, T., et al. 2006, \aap, in press\\
Ryabchikova, T., Sachkov, M., Kochukhov, O., Lyashko, D. 2007, in preparation\\
Sachkov, M., et al. 2004, in {\it Variable Stars in the Local Group}, 
eds. D.W. Kurtz, K.R. Pollard, ASP Conf. Ser., 310, 208\\
Saio, H. 2005, \mnras, 360, 1022\\
Saio, H., Gautschy, A. 2004, \mnras, 350, 485\\
Shibahashi, H., Kurtz, D. W., Kambe, E., Gough, D. O. 2004, in {\it IAU Symposium No. 224, 
The A-star Puzzle}, eds. J.\,Zverko, J.\,\v{Z}i\v{z}\v{n}ovsk\'{y}, \& S.J.\,Adelman, 
W.W.\,Weiss, Cambridge University Press, IAUS~224, 829
}

\end{document}

%% file: CoAst_layo.tex
\renewcommand{\chaptermark}[1]{\markboth{#1}{}}

\newcommand{\apj}{ApJ}
\newcommand{\aap}{A\&A}  %*************Stimmts?
\newcommand{\mnras}{MNRAS}

\def\kms {km~s$^{-1}$}

\newcommand{\kopf}{\small\itshape Comm. in Asteroseismology\\ Vol. 148, 2006}
\newcommand{\Authors}[1]{\begin{center}\normalsize\bf\sf #1 \end{center}}

\renewcommand{\author}[1]{\begin{center}\normalsize\bf\sf #1 \end{center}}
\newcommand{\Address}[1]{\begin{center}\small\sf #1 \end{center}}

\renewenvironment{abstract}{\section*{Abstract}\normalsize\sf}{}
\newcommand{\References}[1]{\begin{flushleft}{\large References\\}\vspace*{2mm}\small #1 \end{flushleft}}

\newcommand{\chapterDSSN}[2]{\chapter[\sf\normalsize #1\\ \footnotesize \hspace*{5mm}by #2 \sf\normalsize][]{#1\\}\rhead[\fancyplain{}{\sf\footnotesize \center{#1}}]{\fancyplain{}{\sffamily\thepage}}\lhead[\fancyplain{\kopf}{\sffamily\thepage}]{\fancyplain{\kopf}{\sf\footnotesize \center{#2}}}}

\renewcommand{\chaptername}{}
\newcommand{\acknowledgments}[1]{\vspace*{5mm}\noindent\begin{bf}Acknowledgments. \end{bf} #1}